# Wideband Coplanar Waveguide MIMO Antenna for 6G Millimeter-Wave Applications with Defected Ground Structure

Atta Ullah[1], Daniyal Munir[1], Daniel Lindenschmitt[1] and Hans D. Schotten[1]
[1]Institute for Wireless Communication and Navigation, RPTU Kaiserslautern-Landau, Kaiserslautern, 67663, Rhineland-Palatinate, Germany
[{atta.ullah, daniyal.munir, daniel.lindenschmit}@rptu.de, schotten@eit.uni-kl.de]

*Abstract*— This research study introduces a novel small antenna with wideband capacity for the higher frequency range. As a possible contender for 6G wireless networks, the proposed antenna is designed to target the 6G Millimeter-Wave (mmWave) operating bands spanning 25 GHz to 33.5 GHz. With a microstrip patch structure fed by a coplanar waveguide (CPW) with the defected ground structure (DGS), a single antenna is introduced and then a design of 2 x 2 MIMO antenna is presented. The single antenna has 2 elements while the 2 x 2 MIMO antenna has 8 elements. It achieves remarkably well in terms of return loss of 8.5 GHz wideband, which is anticipated to be used for several applications in 6G mmWave technology.

## I. INTRODUCTION

The need for wireless machines and their purposes has grown swiftly in the last decade. This covers wireless personal area networks (WPAN), wireless local area networks (WLAN), wireless high-definition multimedia interference (HDMI), peer-to-peer machine communication, and high-speed internet access. The Internet of Things (IoT) and mobile video streaming applications are the main causes of this significant increase in wireless communication traffic [1, 2]. Almost all of these applications require highly interconnected networks of data, people, and other things that can process massive volumes of data in real-time with incredibly low latency. The best way to handle wireless technology's expansion and bandwidth, as well as the growing need for wireless communication systems, is through upcoming FR2 6G millimeter-wave (mmWave) communication which operates at frequencies between 24.25 GHz to 40 GHz [3]. This frequency range has piqued the interest of researchers since it offers a largely unused bandwidth, which is highly needed for future communications systems. For point-to-point and short-distance applications, mmWave offers incredibly rich and potentially directed communication. Moreover, mmWave can also meet the growing need for wireless communications in terms of private and limited communications [4].

Antennas are always associated with wireless communication systems since they are devices that can transform energy or signals in the guiding medium into free space. The late 1970s saw the initial proposal of microstrip antennas for mmWave communications. Microstrip has several benefits, including low cost, thin profile, small volume, ease of integration, and lightweight, which make microstrip patch antennas a viable option for communication systems operating in mmWave frequencies. The sturdy and compact designs of microstrip patch antennas make them an excellent choice for connecting to an integrated circuit in communication systems, specifically for wireless communications. However, the limited bandwidth of the microstrip antenna is a disadvantage, necessitating the use of suitable practices to expand the bandwidth [5]. One of the upcoming advancements for communication devices is anticipated to be an antenna with a broad operating frequency. The feed placement, feeding method, dielectric constant, and substrate depth all distress a microstrip antenna's performance features. Conventional microstrip antennas have several drawbacks, including a single working frequency, poor bandwidth impedance, low gain, a bigger size, and polarization issues. Numerous methods, including stacking, different feeding strategies, frequency selective surfaces (FSS), electromagnetic band gap (EBG), photonic band gap (PBG), metamaterial, etc., have been described to enhance the characteristics of traditional microstrip antennas [6].

A variety of techniques have been employed to attain the intended operating range. One idea that has gained popularity is defected ground structure (DGS), in which the antenna's ground plane is purposefully altered (curtailed or defected) to improve performance. Due to its resonant and band-stop behavior, DGS is frequently employed in microwave resonators, oscillators, phase shifters, filters, individual microstrip antennas, and arrays [7].

A wideband operating frequency can be achieved by the coplanar waveguide (CPW) antenna. In this type of design antenna, the ground plane percentage is designed next to the feeding line of the antenna. A number of CPW slot antenna modifications, including triangle slots, square slots, bow-tie slots and hybrid slots are suggested to boost bandwidth [8].

## II. SINGLE ANTENNA

### A. Antenna Design

The design was made using CST Microwave Studio electromagnetic modelling software. This Fractal-shaped antenna features an exceptionally compact antenna elements on the top of the substrate. The ground plane is removed on two sides and connected via a narrow link at the bottom of the substrate. The proposed antenna system is designed on FR-4 substrate with relative permittivity $\varepsilon_r$ = 4.4 and dielectric loss tangent δ=0.019 with a substrate dimension of 10 x 10 mm$^2$ and a thickness of the substrate is 0.8mm. The antenna patch dimension is 4 x 4 mm$^2$ while the feed line is CPW. On the back side of the substrate, the DGS technique is used to get the required result. The single antenna has two elements. Figure 1 shows the antenna geometry model while Figure 2a display the front side of the antenna and Figure 2b shows the bottom side of the antenna with DGS connected to each other. Table *1* gives the optimized dimensions of the proposed antenna. A radiating patch is fed 50-ohm strip lines to attain specified polarization.

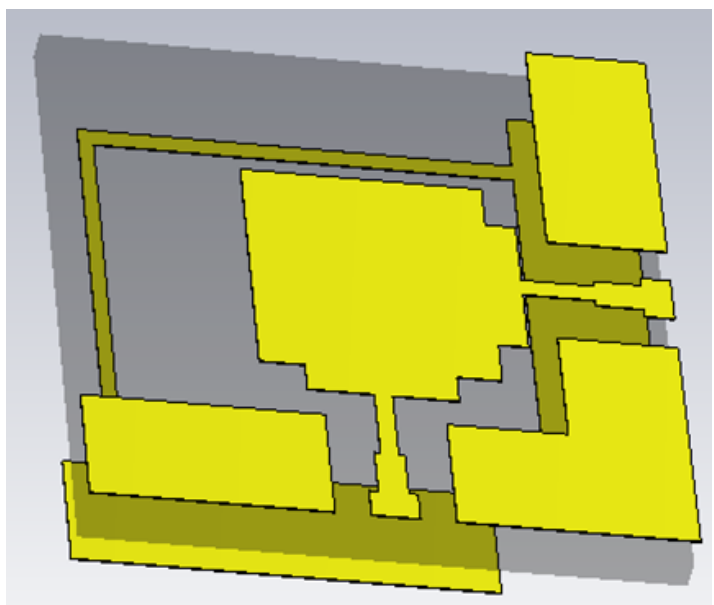

Figure 1: Aerial view of the 2 x 2 MIMO Antenna

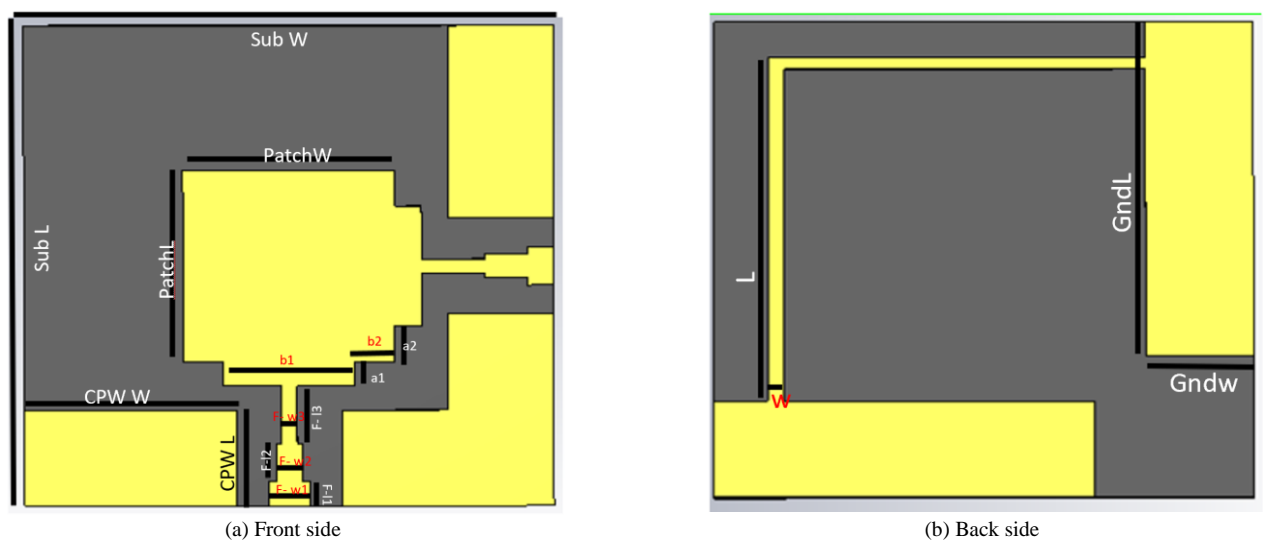

(a) Front side  (b) Back side

Figure 2: Proposed Design of the 2 x 2 MIMO Antenna

Table 1: Dimension of the Proposed Antenna

| Antenna Parameters | Dim. | Antenna Parameters | Dim. | Antenna Parameters | Dim. | Antenna Parameters | Dim. | Antenna Parameters | Dim. |
|---|---|---|---|---|---|---|---|---|---|
| Sub h | 10mm | Sub L | 10mm | Patch W | 4mm | Patch L | 4mm | CPW W | 4mm |
| CPW L | 2mm | F-W1 | 0.8mm | F-W2 | 0.5mm | F-W3 | 0.3mm | F-l1 | 0.5mm |
| F-l2 | 0.8mm | F-l3 | 1.2mm | a1 | 0.5mm | a2 | 0.75mm | b1 | 2.5mm |
| a1 | 0.5mm | a2 | 0.75mm | b1 | 2.5mm | | | | |

## B. Return Loss Results

Figure 3 simulates the return loss of the suggested antenna between 24.5 to 33 GHz. If -10 dB is a reasonable value, this indicates a workable bandwidth of 8.5 GHz. The antenna geometry exhibits resonances at 27.5 GHz and 29.5 GHz. Between 26 and 31 GHz, the design's return loss is good (<-20 dB). Figure 4 shows the mutual coupling between the two elements of the single antenna.

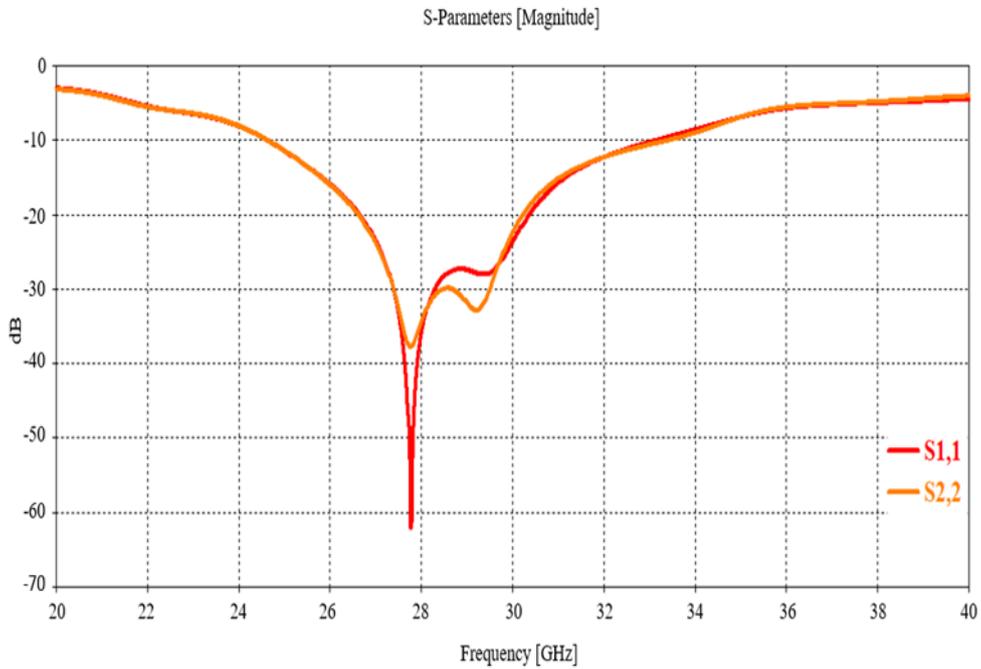

Figure 3: Return loss of the proposed Antenna

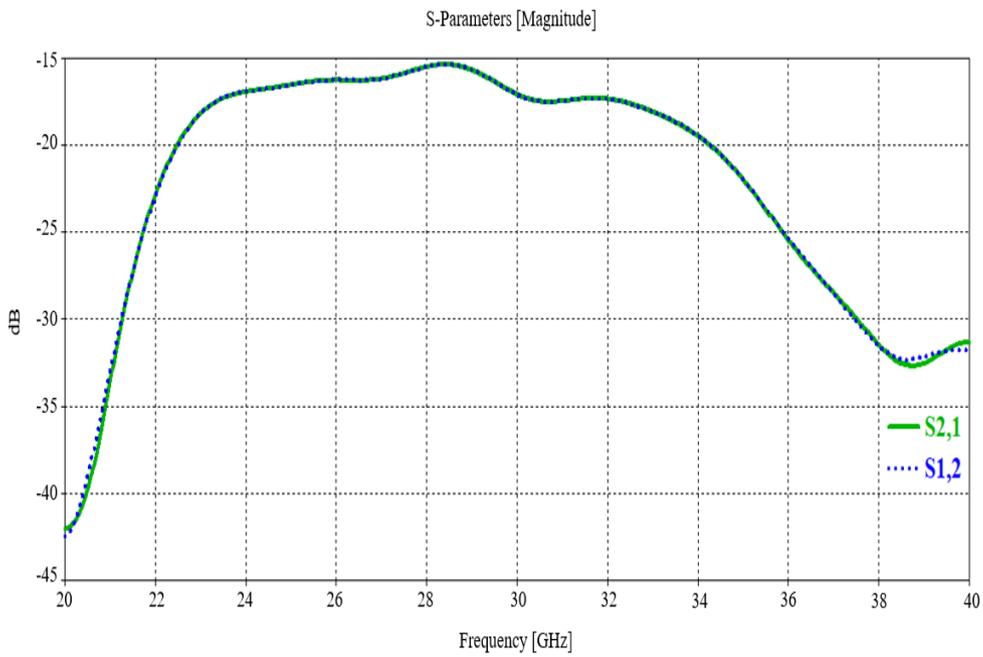

Figure 4: Mutual Coupling of the proposed Antenna

## III. 2 X 2 MIMO ANTENNA

*A. Antenna Design*

This 2 x 2 MIMO Fractal-shape antenna system is incredibly small and is mounted on top of the substrate. It is grounded on two sides and attached to the substrate at the bottom via a tiny connection. The suggested 2 x 2 MIMO antenna system is built on FR-4 substrate, which has dimensions of 20 x 20 mm² and a thickness of 0.8 mm. The relative permittivity is $\varepsilon_r = 4.4$, and the dielectric loss tangent is $\delta = 0.019$. The 2 x 2 MIMO design antenna has eight Elements. The 2 x 2 MIMO simulation-modelled antenna architecture is presented in Figure 5. Figure 6a displays the antenna's upper face, and Figure 6b displays the antenna's bottom side with the DGS.

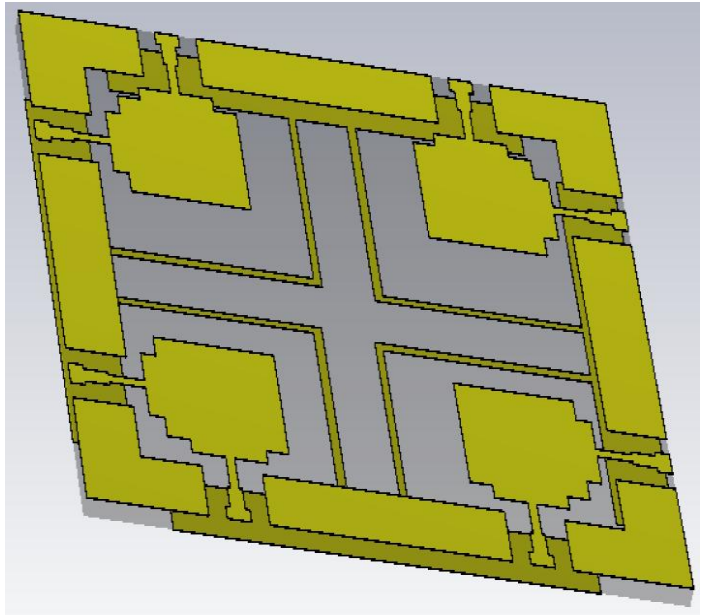

Figure 5: Aerial view of the 2 x 2 MIMO Antenna

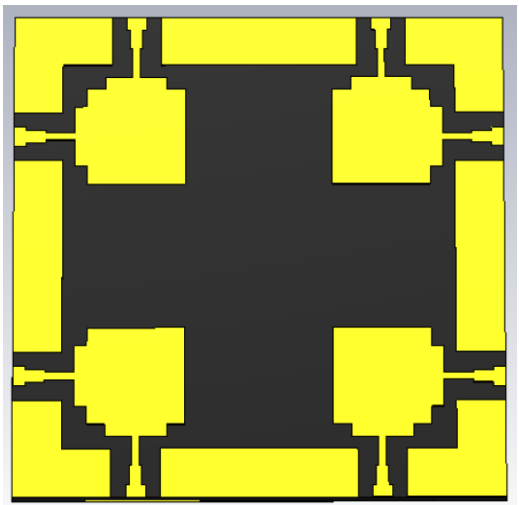 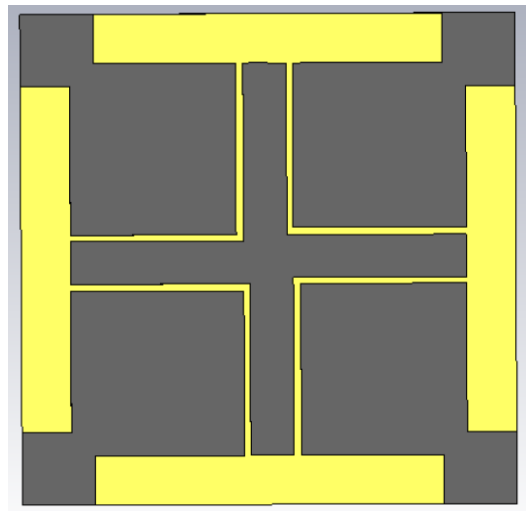

(a) Front side  (b) Back side

Figure 6: Proposed Design of the 2 x 2 MIMO Antenna

## B. Result and discussion of 2x 2 MIMO System

Return loss, resonant frequency, radiation pattern, efficiencies and gain are the constraints which were examined.

### 1) Return Loss Results

Figure 7 simulates the return loss of the proposed 2 x 2 MIMO antenna between 25 GHz to 33.5 GHz. If -10dB is a reasonable value, this indicates a workable bandwidth of 8.5 GHz. The antenna geometry exhibits resonance at 30 GHz. Between 26 and 32 GHz, the design's return loss is good (<-15dB). Figure 8 shows the mutual coupling between the eight elements of the 2 x 2 MIMO antenna system.

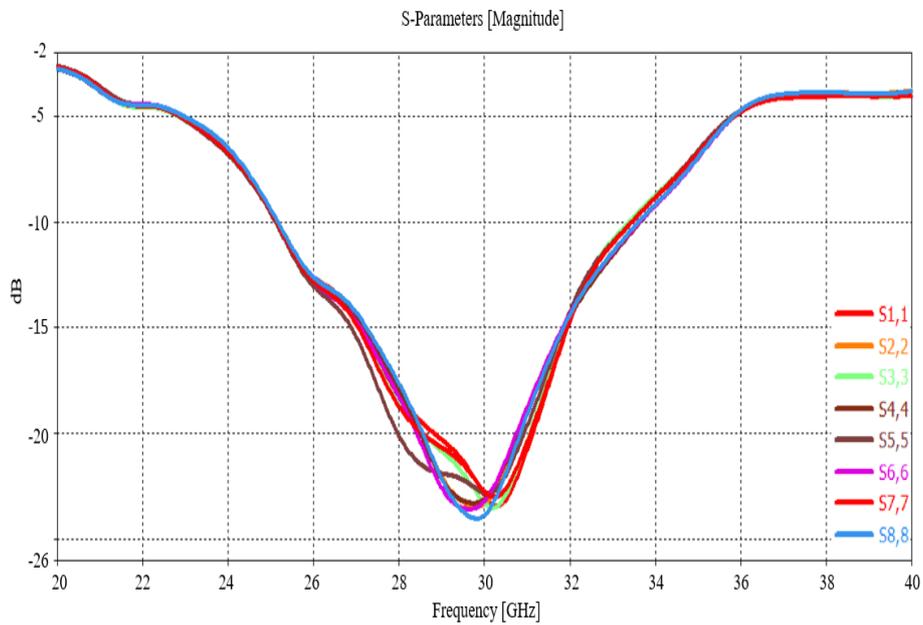

Figure 7: Return loss of the proposed Antenna

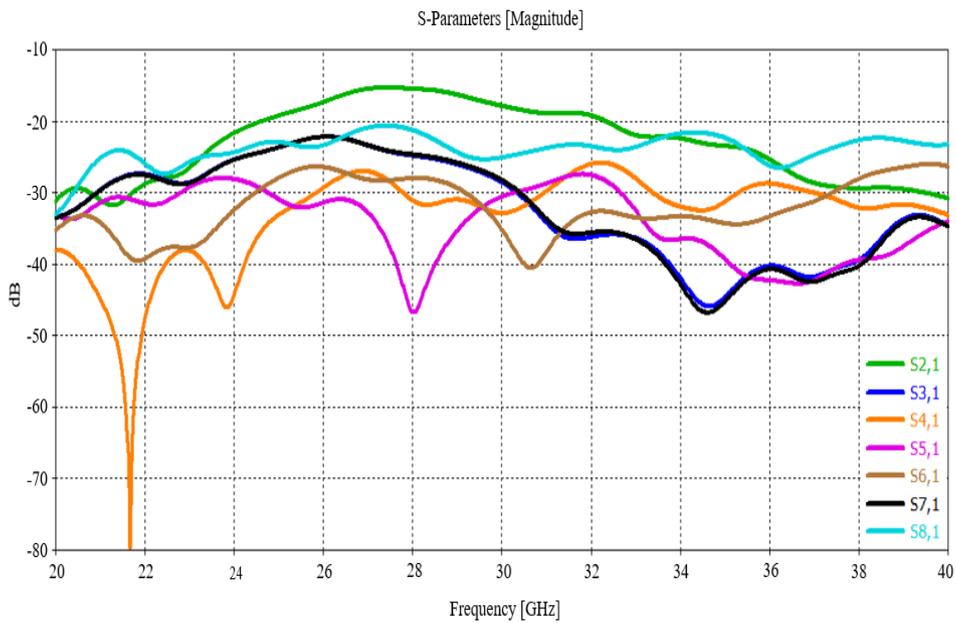

Figure 8: Mutual Coupling of the proposed Antenna

*2) Radiation Performance*

A standardized solid omnidirectional radiation of the designed antenna has been simulated at CST. Figure 9 displays the 3D radiation patterns at 28 GHz. This exhibits a general capacity to imitate omnidirectional patterns, despite considerable unevenness. Figure 10 displays the simulated radiation patterns at 28 GHz that were calculated using principal-plane 2D cuts.

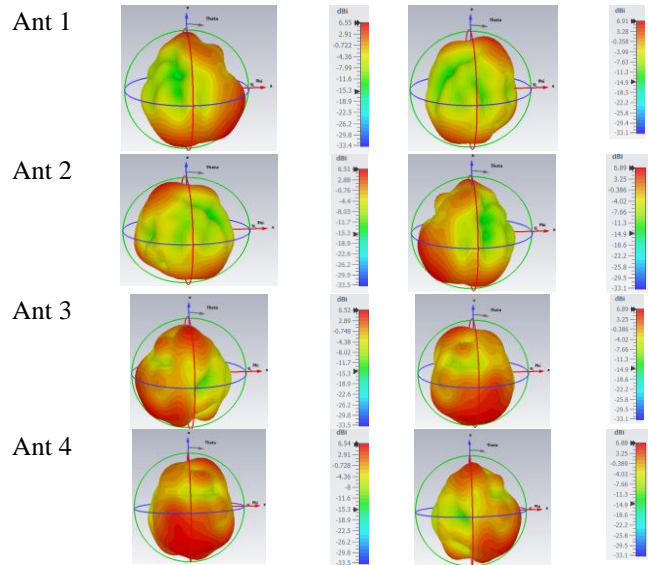

Figure 9: 3D radiation patterns at 28 GHz

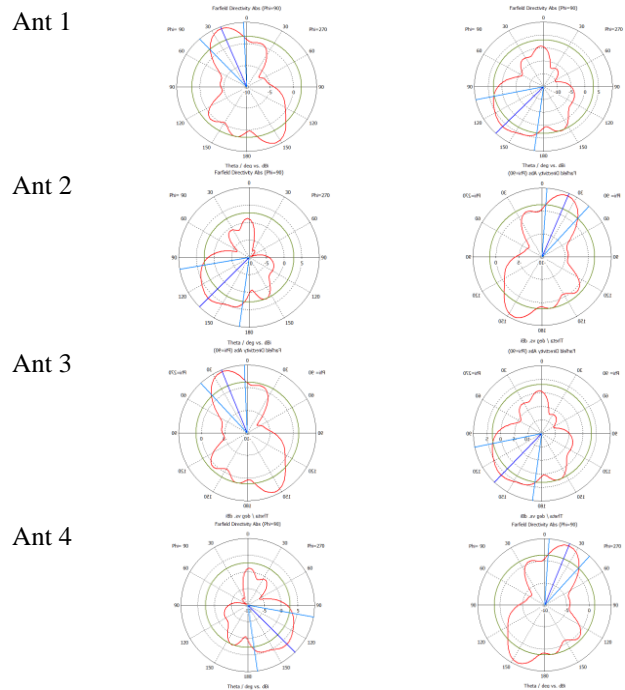

Figure 10: 2D Patterns of proposed Design at 28 GHz

*3) Efficiency*

The radiation and overall efficiency of the suggested CPW-fed antenna are shown in Figure 11. The antenna offers good efficiency over its operating band. It has been observed that more than 60% efficiency has been achieved by the proposed design.

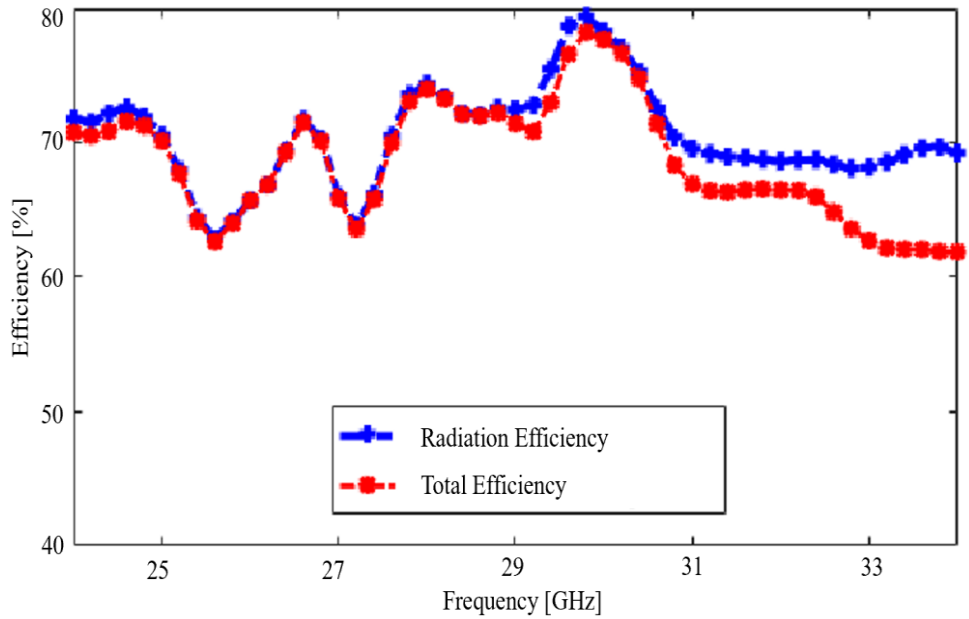

Figure 11: Total and Radiation efficiencies versus frequency of the Antenna

*4) Gain*

Figure 12 displays the antenna's maximum gain in relation to frequency. It is noted that the wideband gain is above 5 dBi between 28.5 and 33 GHz, while the peak gain is above 7 dBi at 32 GHz. However, between 24 GHz and 28 GHz, it is less than 5 dBi.

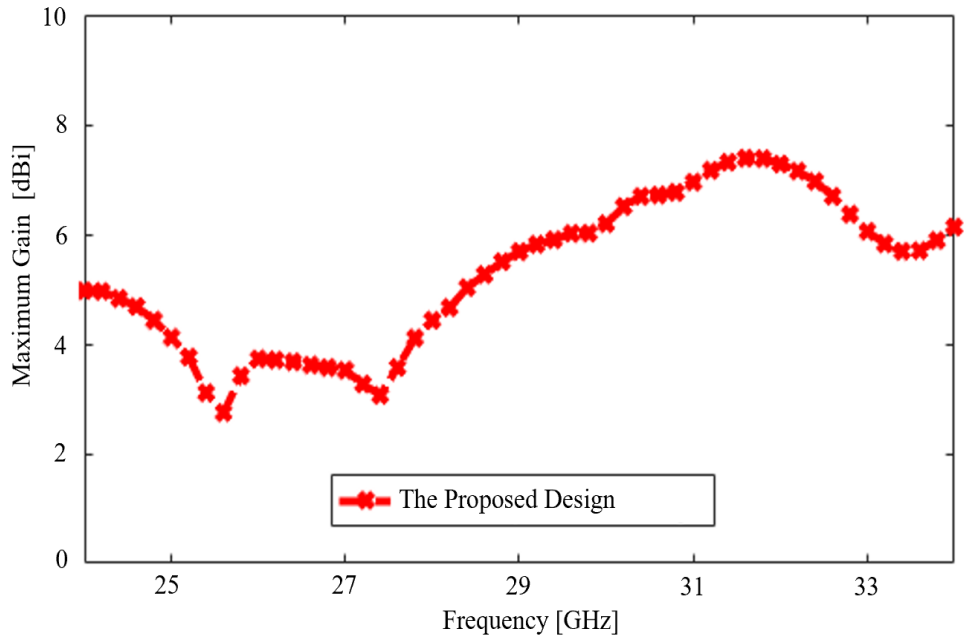

Figure 12: Maximum gain (dBi) vs frequency of the designed antenna

## IV. COMPARISION

In Table 2, the essential features of the suggested antenna are compared to those of the most current 5G antennas that have been published in the literature. In comparison to previously published designs, the suggested antenna may handle a wider impedance bandwidth with greater gain levels, as the table illustrates [9-12].

Table 2: A Comparison of the recently reported and proposed mm wave Antenna

| Reference | Size (mm$^2$) | Operation Band (GHz) | Max Gain (dB) | Isolation (dB) |
|---|---|---|---|---|
| [9] | 20 x 20 | 28/38 | 5.7-7.2 | -16 / -18 |
| [10] | 20 x 20 | 28 / 38 | 1.83 | -20 |
| [11] | 43 x 15 | 28 | N/A | -40 / -38 |
| [12] | 26 x 11 | 27 / 39 | 5-6 | -30 |
| Proposed Antenna | 20 x 20 | 25-33 | 7.5 | <-50 |

## V. CONCLUSION

A 2 x 2 MIMO antenna design has been proposed for upcoming wireless network systems using the 6G mmWave spectrum. It is electrically compact, has a roughly omnidirectional coverage, and is designed to be cost-effective for mass manufacture in commoditised wireless devices. The coplanar waveguide structure on the top surface and the defected ground plane underneath were used to model the design as it would be etched on a high-performance planar substrate. The design had outstanding return loss performance in the 28 GHz band, but it also exhibited a potentially acceptable return loss figure over the broadband from 25 GHz to 33.5 GHz and an appropriate return loss (<-15dB) from 27 to 32 GHz.
.

## ACKNOWLEDGMENT

"The authors acknowledge the financial support by the German Federal Ministry for Education and Research (BMBF) within the projects »Open6GHub« {16KISK004} and »6G-CampuSens« {16KISK208}"